\documentclass[%
 aip,
 sd,%
 amsmath,amssymb,
 reprint,%
]{revtex4-1}
\usepackage{amsmath}
\usepackage{etoolbox}
\usepackage{breqn}
\usepackage{graphicx}% Include figure files
\usepackage{subfigure}
\usepackage{dcolumn}% Align table columns on decimal point
\usepackage{bm}% bold math
\usepackage{float}
\usepackage{hyperref}
\makeatletter
\patchcmd\eq@setnumber{\stepcounter}{\refstepcounter}{}{%
  \errmessage{Patching \noexpand\eq@setnumber failed}%
}
\makeatother
\makeatletter
\let\cat@comma@active\@empty
\makeatother

\begin{document}

\title{Modeling the Network Dynamics of Pulse-Coupled Neurons}

\author{Sarthak Chandra}
\affiliation{University of Maryland, College Park, Maryland 20742, U.S.A.}
\author{David Hathcock}
\affiliation{Case Western Reserve University, Cleveland, Ohio 44016, U.S.A.}
\author{Kimberly Crain}
\affiliation{Iowa State University, Ames, Iowa 50011, U.S.A.}
\author{Thomas M. Antonsen}
\affiliation{University of Maryland, College Park, Maryland 20742, U.S.A.}
\author{Michelle Girvan}
\affiliation{University of Maryland, College Park, Maryland 20742, U.S.A.}
\author{Edward Ott}
\affiliation{University of Maryland, College Park, Maryland 20742, U.S.A.}

\begin{abstract}
We derive a mean-field approximation for the macroscopic dynamics of large networks of pulse-coupled theta neurons in order to study the effects of different network degree distributions, as well as degree correlations (assortativity). Using the ansatz of Ott and Antonsen (\textit{Chaos}, \textbf{19} (2008) 037113), we obtain a reduced system of ordinary differential equations describing the mean-field dynamics, with significantly lower dimensionality compared with the complete set of dynamical equations for the system. We find that, for sufficiently large networks and degrees, the dynamical behavior of the reduced system agrees well with that of the full network. This dimensional reduction allows for an efficient characterization of system phase transitions and attractors. For networks with tightly peaked degree distributions, the macroscopic behavior closely resembles that of fully connected networks previously studied by others. In contrast, networks with scale-free degree distributions exhibit different macroscopic dynamics due to the emergence of degree dependent behavior of different oscillators. For nonassortative networks (i.e., networks without degree correlations) we observe the presence of a synchronously firing phase that can be suppressed by the presence of either assortativity or disassortativity in the network. We show that the results derived here can be used to analyze the effects of network topology on macroscopic behavior in neuronal networks in a computationally efficient fashion.
\end{abstract}

\maketitle
\begin{quotation}
In April 2013, the U.S. President announced `The Brain Initiative,' an extensive, long range plan of scientific research on human brain function. Computer modeling of brain neural dynamics is an important component of this long-term overall effort. A barrier to such modeling is the practical limit on computer resources given the enormous number of neurons in the human brain ($\sim 10^{11}$). Our work addresses this problem by developing a method for obtaining low dimensional macroscopic descriptions for functional groups consisting of many neurons. Specifically, we formulate a mean-field approximation to investigate macroscopic network effects on the dynamics of large systems of pulse-coupled neurons and use the ansatz of Ott and Antonsen to derive a reduced system of ordinary differential equations describing the dynamics. We find that solutions of the reduced system agree with those of the full network. This dimensional reduction allows for more efficient characterization of system phase transitions and attractors. Our results show the utility of these dimensional reduction techniques for analyzing the effects of network topology on macroscopic behavior in neuronal networks.
\end{quotation}

\section{Introduction}

Networks of coupled oscillators have been shown to have a wide variety of biological, physical and engineering applications\cite{michaels1987mechanisms, wiesenfeld1998frequency,kiss2002emerging,motter2013spontaneous, carreras2004complex, glass1973logical, aldana2003natural, luke2013complete, abdulrehem2009low, montbrio2015macroscopic, pazo2014low, laing2014derivation, lu2016resynchronization}. In modelling the dynamics of such networks, simulating the microscopic behavior at each node can be a computationally intensive task, especially when the network is extremely large. In this regard, we note that the dimension reduction analyses in Refs.\cite{ott2008low, ott2009long, ott2011comment} has recently proved to be very effective and has been used to derive the macroscopic behavior of large systems of coupled dynamical units in a variety of settings\cite{restrepo2014mean, montbrio2015macroscopic, martens2009exact, barlev2011dynamics, skardal2015frequency, pazo2014low, pazo2016quasiperiodic, roulet2016average}. In particular, Refs. \cite{luke2013complete, montbrio2015macroscopic, pazo2014low, laing2014derivation} consider network with globally coupled neurons and use these dimension reduction techniques to analyze the macroscopic behavior of the systems.

In 1986, Ermentrout and Kopell\cite{ermentrout1986parabolic} introduced the theta neuron model. Their work, along with later studies by Ermentrout\cite{ermentrout1996type} and by Izhikevich\cite{izhikevich1999class}, established the applicability of the theta neuron model for studying networks of Class I excitable neurons (as defined by Hodgkin,\cite{hodgkin1948local} i.e., those neurons whose activity lies near the transition between a resting state and a state of periodic spiking, and can exhibit spiking with arbitrarily low frequencies).

Previous studies modeling networks of theta neurons\cite{luke2013complete, roulet2016average, martens2009exact, borgers2003synchronization} have generally been restricted to particular classes of network topologies. In this paper we study the macroscopic dynamics of networks of pulse coupled theta neurons on networks with fairly general topologies including arbitrary degree distributions and correlations between the degrees of nodes at 	opposite ends of a link, resulting in so-called `assortativity' or `disassortativity'\cite{newman2002assortative}. Assortativity (disassortativity) occurs when network nodes connect preferentially to those with similar (different) degrees. We note that, studies\cite{hagmann2008mapping, bialonski2013assortative, barzegaran2012properties, de2009functional, teller2014emergence} have shown the biological relevance of assortativity. Motivated by the results of Restrepo and Ott\cite{restrepo2014mean} on networks of Kuramoto oscillators, we use a mean field approach in conjunction with the analytical techniques developed by Ott and Antonsen\cite{ott2008low, ott2009long, ott2011comment} to study the behavior of pulse coupled theta neurons on networks with arbitrary degree distributions and assortativity. We obtain a reduced system of equations describing the mean-field dynamics of the system, with lower dimensionality compared with the complete set of dynamical equations for the system. This allows us to examine the behavior of the network under various conditions in a computationally efficient fashion. We primarily use the example of a scale free degree distribution as an application of the obtained dynamical equations for the order parameter and observe the existence of a partially resting phase, an asynchronously firing phase, and a synchronously firing phase that is sensitive to the presence of assortativity or disassortativity in the network. We also demonstrate that, in contrast to networks with sharply peaked degree distributions, networks with scale-free degree distributions exhibit different macroscopic dynamics due to the emergence of degree dependent behavior of different oscillators. 

The remainder of this paper is organized as follows. In Sec. \ref{sec:model} we describe the model of pulse coupled theta neurons used on an arbitrary network. In Sec. \ref{sec:meanfield} setup a mean field description of the behavior on the network, and then (Sec. \ref{sec:dimensionreduction}) show how the methods developed by Ott and Antonsen\cite{ott2008low, ott2009long, ott2011comment} can be used to write a low dimensional set of equations describing the dynamics of the mean field order parameter. In Sec. \ref{sec:numerics} we then use this low dimensional system to describe the behavior of the system under different parameters and network topologies. Section \ref{sec:conclusion} concludes the paper with further discussion and summary of the main result.

\section{The model}\label{sec:model}

\begin{figure}
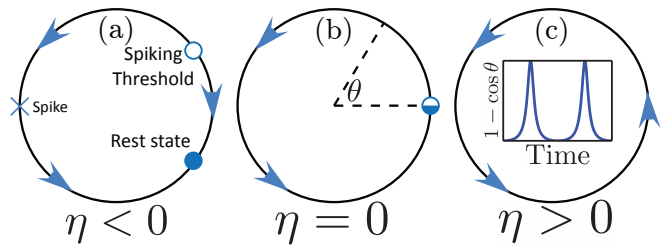

\includegraphics[width=\columnwidth]{{{Theta_neuron_a_b_c}}}
\caption{The dynamics of the theta neuron undergo an SNIC (Saddle node on an Invariant Cycle) bifurcation at $\eta=0$. For negative $\eta$ the neuron lies in a rest state, with a threshold for excitation, and for positive $\eta$ the oscillator undergoes periodic spiking.}
\label{fig:thetaneuron}
\end{figure}

The theta neuron model encodes the dynamics of a single neuron in isolation as follows,
\begin{equation}\label{eq:thetaneuron}
\dot \theta = (1-\cos\theta) + (1+\cos\theta)\eta,
\end{equation}
where $\theta$ represents the neuron's state and the parameter $\eta$ specifies its excitability. The dynamics can be visualized as a point traveling around the unit circle (Fig. \ref{fig:thetaneuron}). A neuronal spike is said to occur each time the phase angle of the neuron, $\theta$, crosses the leftmost point at $\theta=\pi$. When $\eta <0$, there are two zeros of the right hand side of Eq. (\ref{eq:thetaneuron}), representing a stable rest state (solid circle in Fig. \ref{fig:thetaneuron}(a)) and an unstable equilibrium (open circle in Fig. \ref{fig:thetaneuron}(a)). Thus, starting from a typical initial condition, the state of the neuron goes towards the stable equilibrium at the rest state represented by the filled circle. A resting neuron will spike if an external force pushes its state (i.e. the angle $\theta$) from the rest state past the unstable equilibrium (termed as the `spiking threshold'). As $\eta$ is increased above $0$, the neuron exhibits a Saddle Node bifurcation on an Invariant Cycle (SNIC). In this case there are no fixed points (i.e. no zeros of the right hand side of Eq. (\ref{eq:thetaneuron})), and the neuron now fires periodically, as shown in Fig. \ref{fig:thetaneuron}(c). Note that the neuron does not move at the same rate along the entire circle, and may go faster or slower around $\theta=\pi$ dependent on whether $\eta$ is less than or greater than $1$, respectively (eq. see the plot of $(1-\cos\theta)$ versus time in Fig. \ref{fig:thetaneuron}(c)).

The theta neuron model can be extended from a single neuron in isolation to networks of neurons. We consider a system of $N$ theta neurons coupled together in a general network via pulse-like synaptic signals, $I_i$, to each neuron $i$:

\begin{align}\label{eq:network}
\dot \theta_i &= (1- \cos \theta_i) + (1+ \cos\theta_i)[\eta_i+ I_i],\\
I_i &= \frac{K}{\langle k \rangle} \sum_{j=1}^N A_{ij} P_n(\theta_j),
\end{align}
where $A_{ij}$ is the adjacency matrix of a network; $A_{ij}=1$ if there is a directed edge from node $j$ to node $i$, and $A_{ij}=0$ otherwise. The average degree is then given by $\langle k \rangle = \sum_{i,j} A_{ij}/N$. $P_n(\theta) = d_n(1-\cos\theta)^n$ represents the pulse-like synapse, whose sharpness is controlled by the integer parameter $n$. The normalization constant $d_n$ is determined so that $\int_0^{2\pi} P_n d\theta = 2 \pi $. Note that in the case of a fully connected network, where $A_{ij} =  1$ for all $i$ and $j$, this model reduces to that of Luke et al.\cite{luke2013complete}

\section{Mean Field Formulation}\label{sec:meanfield}

We consider the limit of many neurons, $N \gg 1$, and assume the network is randomly generated from a given degree distribution $P(\mathbf{k})$ (normalized such that $\sum_{\mathbf{k}} P(\mathbf{k}) = N$), where  $\mathbf{k}$, the node degree, represents a two-vector of the in-degree and the out-degree, $(k_{in}, k_{out})$. Additionally, we consider an assortativity function $a(\mathbf{k'} \rightarrow \mathbf{k})$, which specifies the probability of a link from a node of degree $\mathbf{k'}$ to one of degree $\mathbf{k}$. In this $N\to\infty$ limit, we assume that the state of the neurons can be represented by a continuous probability distribution, $f(\theta, \eta | \mathbf{k}, t)$, such that $f(\theta, \eta | \mathbf{k}, t)d\theta d\eta$ is the probability that a node of degree $\mathbf{k}$ has an excitability parameter in the range $[\eta, \eta+d\eta]$ and a phase angle in the range $[\theta, \theta + d\theta]$ at time $t$. Since we are assuming that the excitability parameters do not vary with time, we define $g(\eta | \mathbf{k}) = \int f d\theta$, which is the time independent distribution of the excitability parameters $\eta_i$ in the network for a randomly chosen node of degree $\mathbf{k}$.

In order to describe the synchronization behavior of this system, we define the order parameter to be\footnote{Some authors, such as Restrepo and Ott\cite{restrepo2014mean} define the order parameter differently so as to be weighted with the out-degree at each node, i.e., $R(t) = \sum_{i=1}^N \sum_{j=1}^N  A_{ij} e^{i \theta_j} / \left(\sum_{i=1}^N \sum_{j=1}^N A_{ij}\right)$}, 

\begin{equation} \label{eq:NetworkOrder}
R(t) = \frac{1}{N} \sum_{j=1}^N e^{i \theta_j}.
\end{equation}
As in previous work by Restrepo and Ott\cite{restrepo2014mean}, we hypothesize that in networks with large nodal degrees, the order parameter can be well approximated via a mean field order parameter, defined by a continuum version of Eq. (\ref{eq:NetworkOrder}),
\begin{equation} \label{eq:MForder}
\bar{R}(t) = \frac{1}{N} \sum_{\mathbf{k'}} P(\mathbf{k'}) \int \int f(\theta', \eta' | \mathbf{k'}, t) e^{i \theta'} d\theta' d\eta'.
\end{equation}
Additionally, the distribution $f$ is constrained by the continuity equation,
\begin{equation} \label{eq:contEq}
\frac{\partial f}{\partial t} + \frac{\partial}{\partial \theta} (v_\theta f) = 0,
\end{equation}
where $v_\theta$ is the continuous version of the right hand side of Eq. (\ref{eq:network}),
\begin{dmath} \label{eq:contV}
v_\theta = (1- \cos \theta) + (1+ \cos \theta) \left[\eta + d_n \frac{K}{\langle k \rangle} \sum_{\mathbf{k'}} P(\mathbf{k'}) a({\mathbf{k'} \rightarrow \mathbf{k}})\\ \times \int \int f(\theta', \eta' | \mathbf{k'}, t)  (1- \cos \theta')^n d\theta' d\eta' \right].
\end{dmath}

\section{Dimension Reduction}\label{sec:dimensionreduction}

Employing the dimensional reduction method of Ott and Antonsen \cite{ott2008low, ott2009long, ott2011comment}, and following its previous application to the theta neuron\cite{luke2013complete}, we assume that $f$ is given by the Fourier expansion,
\begin{dmath} \label{eq:ansatz}
f(\theta, \eta | \mathbf{k}, t) = \frac{g(\eta | \mathbf{k})}{2 \pi} \left\{1\\ + \sum_{p = 1}^\infty \left[b(\eta, \mathbf{k}, t)^p e^{-i p \theta} +  b^*(\eta, \mathbf{k}, t)^p e^{i p \theta} \right] \right\}.
\end{dmath}
We then use the binomial theorem to expand the pulse function $P_n(\theta)$ using
\begin{equation} \label{eq:binomExp}
(1-\cos \theta)^n = A_0 + \sum_{p = 1}^n A_p [e^{i p \theta} + e^{-i p \theta}],
\end{equation}
where
\begin{equation} \label{eq:Acoeff}
A_p = \sum_{j,m=0}^n \delta_{j-2m, p} Q_{jm},
\end{equation}
and
\begin{equation} \label{eq:Qcoeff}
Q_{jm} = \frac{(-1)^j n!}{2^j m! (n-j)! (j-m)!}.
\end{equation}
If we now assume a Lorentz distribution of the excitability parameters, 
\begin{equation} \label{eq:Lorentzian}
g(\eta | \mathbf{k}) = \frac{1}{\pi} \frac{\Delta(\mathbf{k})}{[\eta - \eta_0(\mathbf{k})]^2 + \Delta^2(\mathbf{k})},
\end{equation}
we obtain
\begin{equation} \label{eq:expIntegral}
\int \int f(\theta', \eta' | \mathbf{k}, t) e^{i p \theta} d\theta' d\eta' = \left\{
        \begin{array}{ll}
            \hat b(\mathbf{k}, t)^p, & \quad p > 0 \\
            1, & \quad p = 0 \\
            \hat b^*(\mathbf{k},t)^{|p|}, & \quad p < 0,
        \end{array}
    \right.
\end{equation}
with $\hat b(\mathbf{k}, t) \equiv b(\eta_0(\mathbf{k}) + i \Delta(\mathbf{k}), \mathbf{k}, t)$.
This now allows us to rewrite $v_{\theta}$ in terms of $\hat b(\mathbf{k},t)$ as
\begin{equation} \label{eq:vEq}
v_\theta = g e^{i \theta} + h + g^* e^{-i \theta},
\end{equation}
where
\begin{equation} \label{eq:g&h}
g = -\frac{1}{2} (1 - \eta - \frac{K}{\langle k \rangle} H_n( \mathbf{k}, t)), \quad h = 1 + \eta + \frac{K}{\langle k \rangle} H_n( \mathbf{k}, t),
\end{equation}
and 

\begin{dmath} \label{eq:Hn}
H_n(\mathbf{k}, t) = d_n \sum_{\mathbf{k'}}\left\{ P(\mathbf{k'}) a({\mathbf{k'} \rightarrow \mathbf{k}}) \\ \times \left[ A_0 + \sum_{p=1}^n A_p (\hat b(\mathbf{k'}, t)^p + \hat b^*(\mathbf{k'}, t)^p) \right] \right\}.
\end{dmath}
Substituting the phase velocity Eq. (\ref{eq:vEq}) and the Ott-Antonsen ansatz Eq. (\ref{eq:ansatz}) into the continuity equation (\ref{eq:contEq}), we find that $b(\eta, \mathbf{k}, t)$ satisfies:

\begin{equation} \label{eq:beq}
\frac{\partial b}{\partial t} = i (g b^2 + h b + g^*).
\end{equation}

Inserting the forms for $g$ and $h$ from Eq. (\ref{eq:g&h}) and (\ref{eq:Hn}) into this expression, and evaluating each quantity at the pole, $\eta = \eta_0(\mathbf{k}) + i \Delta(\mathbf{k})$, we obtain a reduced system of equations for $\hat b(\mathbf{k}, t)$ describing the mean field dynamics of the neuronal network, 

\begin{dmath}\label{eq:reducedeqn}
\frac{\partial{\hat b(\mathbf{k}, t)}}{\partial t} = -i \frac{(\hat b(\mathbf{k}, t)-1)^2}{2} + \frac{ (\hat b(\mathbf{k}, t) + 1)^2}{2} \left\{-\Delta(\mathbf{k}) + i \eta_0(\mathbf{k}) + i d_n \frac{K}{\langle k \rangle} \sum_{\mathbf{k'}} P(\mathbf{k'}) a({\mathbf{k'} \rightarrow \mathbf{k}}) \\ \times  \left[ A_0 + \sum_{p=1}^n A_p (\hat b(\mathbf{k'}, t)^p + \hat b^*(\mathbf{k'}, t)^p) \right]\right\}.
\end{dmath}
The mean field order parameter, $\bar{R}(t)$ can now be written in terms of $\hat b(\mathbf{k},t)$. Using the assumed form for $f(\theta,\eta|\mathbf{k},t)$, we can evaluate the integrals in Eq. (\ref{eq:MForder}) using Cauchy's residue theorem to obtain
\begin{equation} \label{eq:MForder2}
\bar{R}(t) = \frac{1}{N} \sum_{\mathbf{k}} P(\mathbf{k}) \hat b(\mathbf{k},t).
\end{equation}

For the discussion in this paper, we will restrict the assortativity function to be of the form used previously by Restrepo and Ott \cite{restrepo2014mean}
\begin{equation} \label{eq:assort1}
a(\mathbf{k'} \rightarrow \mathbf{k}) = h(a_{\mathbf{k'} \rightarrow \mathbf{k}}),
\end{equation}
where $h(x)=\min(\max(x,0),1)$ is defined to ensure that $a(\mathbf{k'} \rightarrow \mathbf{k})$ is a valid probability (i.e. $0\leq a(\mathbf{k'} \rightarrow \mathbf{k}) \leq 1$), and 
\begin{equation}\label{eq:assort}
a_{\mathbf{k'} \rightarrow \mathbf{k}}  = \frac{k_{out}' k_{in}}{N \langle k \rangle} \left[1 + c \left(\frac{k_{in}' - \langle k \rangle}{k_{out}'} \right) \left( \frac{k_{out} - \langle k \rangle}{k_{in}} \right) \right],
\end{equation}
where c is a parameter used to vary the network assortativity (with $c>0$ and $c<0$  corresponding to assortative and disassortative networks, respectively). In networks with neutral assortativity ($c=0$), the probability of forming a link between two nodes is simply proportional to the out-degree of the source node and the in-degree of the target node. 

The in-out Pearson assortativity coefficient, $r$, is a statistic used to characterize the overall assortativity of a network, and is defined\cite{foster2010edge} as 
\begin{equation} \label{eq:pearson}
r = \frac{\sum_e \left[ (k_{in}' - \langle k \rangle) (k_{out} - \langle k \rangle) \right]}{\sqrt{\sum_e (k_{in}' - \langle k \rangle)^2} \sqrt{\sum_e (k_{out} - \langle k \rangle)^2}},
\end{equation}
where $\sum_e$ is the sum over all edges connecting a node of degree $\mathbf{k'}$ to a node of degree $\mathbf{k}$\footnote{For another, often useful, definition of a coefficient quantitatively characterizing the assortativity or disassortativity of a network see Ref.\cite{restrepo2007approximating}}. Assuming that $a(\mathbf{k'} \rightarrow \mathbf{k}) = a_{\mathbf{k'} \rightarrow \mathbf{k}}$, and that the in and out degree distributions are independent, we can relate the assortativity coefficient to the parameter $c$ as 
\begin{equation}\label{eq:pearsonC}
r = \frac{c}{\langle k \rangle^2} \sqrt{(\langle k_{in}^2 \rangle - \langle k \rangle^2)(\langle k_{out}^2 \rangle - \langle k \rangle^2)},
\end{equation}
which can be seen by noting that the sum of a quantity $Q(\mathbf{k},\mathbf{k'})$, defined on each edge connecting a node of degree $\mathbf{k'}$ to a node of degree $\mathbf{k}$, over edges in our mean field formulation would be given by $\sum_e Q(\mathbf{k},\mathbf{k'}) = \sum_{\mathbf{k}} \sum_{\mathbf{k'}} P(\mathbf{k'}) a(\mathbf{k'} \rightarrow \mathbf{k}) P(\mathbf{k}) Q(\mathbf{k},\mathbf{k'})$.

The expression for the assortativity coefficient as a function of $c$, Eq. (\ref{eq:pearsonC}), is unbounded, while the Pearson assortativity is by definition bounded between $-1$ and $1$. This difference arises because, for sufficiently large $c$, the assortativity function given in Eq. (\ref{eq:assort}) is not a probability. However, for the network parameters used in our numerical example below, we find that Eq. (\ref{eq:pearsonC}) is very accurate for $|c| \leq 2.5$, corresponding to an assortativity range, $|r| \lesssim 0.198 $.

If we assume the excitability parameters are drawn from a degree independent distribution ($g(\eta |\mathbf{k}) \equiv g(\eta)$) and the $\hat b$'s are given $\mathbf{k}$ independent identical initial conditions, $\hat b(\mathbf{k}, 0) \equiv \hat b(0)$, then there are a few notable cases in which particular degree distributions and our chosen assortativity function Eq. (\ref{eq:assort}) allow for further dimensional reduction. For networks with a delta-function degree distribution, $P(\mathbf{k}) = \delta_{k_{in}, k} \delta_{k_{out}, k}$, the Eq. (\ref{eq:reducedeqn}) reduces to a single equation describing the mean field dynamics, 

\begin{dmath} \label{eq:singleEqRed}
\frac{\partial{\hat b(t)}}{\partial t} = -i \frac{(\hat b(t)-1)^2}{2} + \frac{ (\hat b(t) + 1)^2}{2} \left\{-\Delta + i \eta_0 + i d_n K \left[ A_0 + \sum_{p=1}^n A_p (\hat b(t)^p + \hat b^*(t)^p) \right]\right\}.
\end{dmath}
We note that this equation is identical to earlier results for a fully connected network\cite{luke2013complete}. Thus, networks with only a single allowed degree have identical asymptotic dynamics to a fully connected network. This result is consistent with analogous results by Barlev et al\cite{barlev2011dynamics} for a network of Kuramoto oscillators. More generally, if the network has fixed in-degree, $P(\mathbf{k}) = P(k_{out}) \delta_{k_{in}, k}$, the system is similarly reduced to the single dynamical equation, Eq. (\ref{eq:singleEqRed}). On the other hand, if the out-degree is fixed,  $P(\mathbf{k}) = P(k_{in}) \delta_{k_{out}, k}$, then dynamics of $\hat b(\mathbf{k},t)$ is independent of $k_{out}$, further reducing the dimensionality of the problem.

\subsection*{Reduction efficiency}
Equation (\ref{eq:reducedeqn}) represents a reduction of the original system of $N$ theta neurons to a system with as many equations as there are \textbf{k} values in the support of the degree distribution $P(\mathbf{k})$. We denote this quantity by $M_{\mathbf{k}}$, which, in the case of independent in and out-degree distributions, is equal to $M_{in} \times M_{out}$, where $M_{in}$ and $M_{out}$ are the number of possible in-degrees and out-degrees respectively. In general, simulating the full network, Eq. (\ref{eq:network}), requires $\mathcal{O}(N^2)$ floating point operations per time step. Using the form of the assortativity function given in Eq. (\ref{eq:assort}) the sum over $\mathbf{k'}$ in the reduced system of equations can be split into two sums, each independent of $\mathbf{k}$,

\begin{equation} \label{eq:splitSum}
\frac{k_{in}}{N \langle k \rangle} \sum_{\mathbf{k'}} P(\mathbf{k'}) k_{out}'  \mathcal{A}  + c \frac{k_{out}-\langle k \rangle}{N \langle k \rangle}  \sum_{\mathbf{k'}} P(\mathbf{k'})  (k_{in}' - \langle k \rangle) \mathcal{A}. 
\end{equation}
where $\mathcal{A} =  A_0 + \sum_{p=1}^n A_p \left(\hat b(\mathbf{k'}, t)^p + \hat b^*(\mathbf{k'}, t)^p\right)$. 
Since the two sums in Eq. (\ref{eq:splitSum}) are independent of $\mathbf{k}$, each must be calculated only once per simulation iteration. Thus, simulating the reduced system Eq. (\ref{eq:reducedeqn}) only requires $\mathcal{O}(M_{\mathbf{k}})$ floating point operations per time step --- $M_{\mathbf{k}}$ operations performed once for each of these two sums and $M_{\mathbf{k}}$ operations for each of the $\hat b(\mathbf{k},t)$ equations. In many cases, $M_{\mathbf{k}} \ll N^2$, so that simulating Eq. (\ref{eq:reducedeqn}) is significantly more efficient that simulating the full network. Furthermore, if $c$ is set to $0$, which is the case of networks with neutral assortativity, then $\hat b(\mathbf{k}, t)$ will have no dependence on $k_{out}$, and hence the overall problem is reduced to $M_{in}$ independent equations, allowing even greater computational efficiency.

\begin{figure}
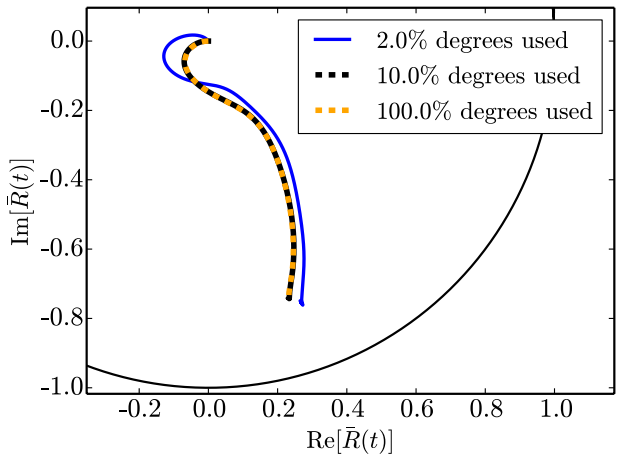

\includegraphics[width=\columnwidth]{{{FixedPoints_meanfield_eta-2_delta_0.1_K3._c0_varying_interpolation_legends_dashed_axes_labels}}}
\caption{The effect of varying levels of interpolation on the calculated results for the trajectories of $\bar{R}(t)$ in the complex plane starting from an initial condition of $\bar{R}(t)=0$ and ending at a fixed point attractor for $K=3$ in a network with neutral assortativity, with $\eta_0=-2$ and $\Delta=0.1$. Calculation of the order parameter dynamics is robust to a large range in the level of interpolation. Using as few as 10\% of the total available degrees and interpolating the remaining 90\% give results close to the calculation without interpolation. In the rest of this paper we employ a 10\% interpolation level in all our mean field calculations. The black arc is a segment of the unit circle $|\bar{R}(t)|=1$.
}\label{fig:interpolation}
\end{figure}

Since $\hat b(\mathbf{k'}, t)$, $P(\mathbf{k'})$, and $a(\mathbf{k'} \rightarrow \mathbf{k})$ are each smoothly varying functions, we can achieve further dimensional reduction by interpolating the summand in Eq. (\ref{eq:reducedeqn}) using a coarse-grained grid of $\mathbf{k}$ values. In particular, Eq. (\ref{eq:reducedeqn}) is not solved for $\hat b(\mathbf{k},t)$ for all of the $M_{\mathbf{k}}$ values of $\mathbf{k}$, but only for the small subset of $\mathbf{k}$ values that lie on the coarse-grained grid in $\mathbf{k}$-space The summands on the right hand side of Eq. (\ref{eq:reducedeqn}) at $\mathbf{k}$ values not on the grid are approximated by a bilinear interpolation of the values at the surrounding chosen $\mathbf{k}$ values. To perform the bilinear interpolation, we first interpolate linearly between neighboring grid values in one direction. The value of the summand at a given $\mathbf{k}$ value is then approximated by linearly interpolating in the other direction between values estimated with the previous linear interpolation. We find that using as few as $10\%$ of the network degrees yield very accurate results, while an even coarser interpolation still produces the same qualitative behavior as can be seen in Fig. \ref{fig:interpolation}.

\section{Numerical simulations and results}\label{sec:numerics}

In the following examples, we consider a directed network of $N=5000$ nodes, with in and out degrees chosen from independent, identical heavy-tailed distributions given by
\begin{equation}\label{eq:degdist}
P(k) = \begin{cases}
0             &\text{if $k<k_{min}$} \\
A k^{-\gamma} &\text{if $k_{min}\leq k < k_{max}$} \\
0             &\text{if $k_{max}\leq k$}.
\end{cases}
\end{equation}
The exponent of the power law distribution, $\gamma$, was set to $3$, and $k_{min}$ and $k_{max}$ were set to $750$ and $2000$, respectively. As mentioned earlier, the normalization constant $A$ is chosen to make $\sum_{\mathbf{k}} P(\mathbf{k}) = N$. We will also set the parameter $n$ controlling the sharpness of the synaptic pulse to $2$ for all examples considered, and will use an interpolation level of 10\% for all calculations using the reduced system of equations for the mean field theory (cf. Fig. \ref{fig:interpolation}).

From numerical simulations of the reduced equations, (\ref{eq:reducedeqn}), we find that the long term dynamics of the order parameter can be broadly classified into one of three phases -- (1) the partially resting (PR) phase; (2) the asynchronously firing (AF) phase; and (3) the synchronously firing (SF) phase. The PR phase and the AF phase appear as fixed points in the dynamics of the order parameter, whereas the SF phase appears as a limit cycle of the order parameter.

\subsection{Fixed points}

\begin{figure*}
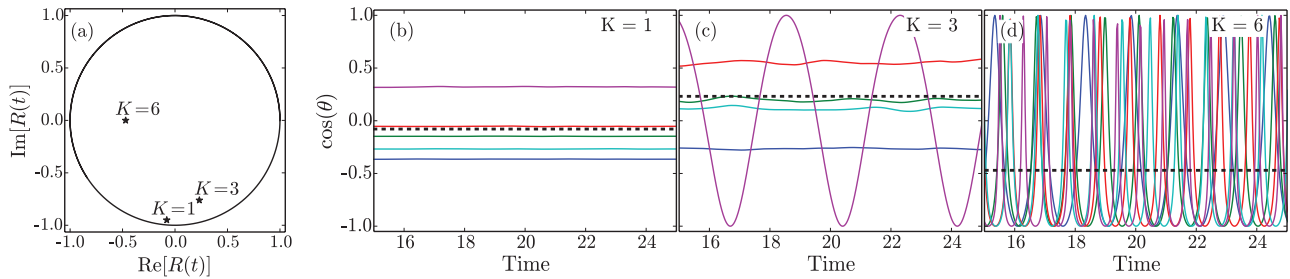

\centering
\includegraphics[width=2.\columnwidth]{{{FixedPoints_time_series_network_eta-2_delta_0.1_a_b_c_d_1}}}
\caption{(a): Fixed points of $R(t)$ observed in networks with neutral assortativity,  $\eta_0=-2$ and $\Delta=0.1$, for three values of the coupling strength $K$. Fixed points in the PR state ($K=1$) and the AF state ($K=6$) are marked in the complex plane. The fixed point at an intermediate value of $K$ is also marked. (b),(c),(d): Time series of the cosine of the phase of 5 randomly chosen neurons demonstrates that in the PR phase almost all neurons are in a resting state, and as the system approaches the AF state, more nodes transition to an oscillating, excited state. The thick dashed line corresponds to the position of the fixed point of the order parameter for the corresponding value of $K$.
}\label{fig:fixedpoints}
\end{figure*}

As a particular example to illustrate the different types of fixed points, we look at a network with neutral assortativity ($c=0$) having excitability parameters distributed according to a Lorentzian distribution with mean $\eta_0=-2$ and width $\Delta=0.1$ (Fig. \ref{fig:fixedpoints}).

When the network is in the PR phase, the order parameter goes to a fixed point that lies near the edge of the unit circle $|\bar{R}|=1$. In this phase, most of the individual neurons in the network are independently in their resting states, in a fashion similar to Fig. \ref{fig:thetaneuron}(a). This corresponds to the case of $K=1$ in Fig. \ref{fig:fixedpoints}(a), in which the fixed point is located near the edge of the unit circle marked in black. Further, the time series of a few randomly chosen neurons (Fig. \ref{fig:fixedpoints}(b)) demonstrates that almost all of the neurons are in a resting state. While there may be a small number of neurons that are in the spiking phase due to the spread in the distribution of values of excitability parameters, $\eta$, these do not have any significant effect on the full order parameter of the system.

As we increase the coupling constant $K$, the system transitions to the asynchronously firing (AF) phase, in which the order parameter goes to a fixed point located near the center of the unit circle. In this phase, most of the individual neurons in the network are asynchronously firing, in a fashion similar to  Fig. \ref{fig:thetaneuron}(c). This can be seen in the case of $K=6$ in Fig. \ref{fig:fixedpoints}(c) which shows that almost all of the neurons are in a recurrent spiking state. Note that even though the neurons are spiking \emph{asynchronously}, i.e., their firing times are independent of one another\footnote{this definition of asynchronous spiking is consistent with remarks by other authors \cite{abbott1993asynchronous, hansel2001existence}, wherein asynchronous states have been defined as states in which at each neuron the term coupling it to the other neurons in the network is independent of time, as is observed in the cases of fixed points.}, the fixed point of the order parameter is not at $\bar{R}=0$. This is because the angular velocity of an individual neuron is not constant along the circle, thus in the average over the ensemble of neurons a bias is present towards the direction for which the angular velocity of neurons is minimized. As discussed in Sec. \ref{sec:model}, this may occur at either $\theta=0$ or at $\theta=\pi$, dependent on how large the excitability parameter is for the neuron.

\begin{figure}
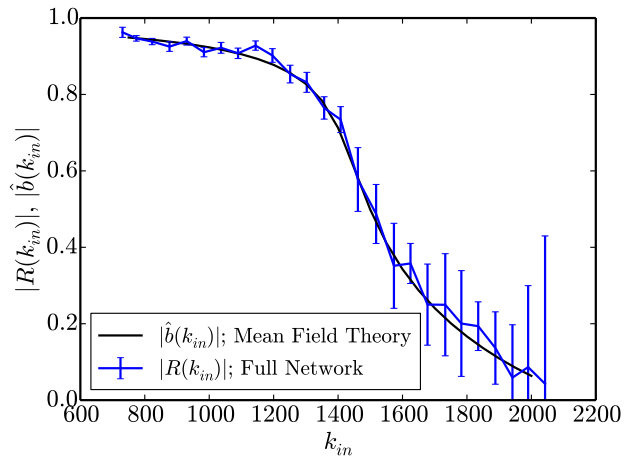

\includegraphics[width=\columnwidth]{{{FixedPoints_meanfield_eta-2_delta_0.1_K3._degree_variation_modulus_bigger_correct_labels}}}
\caption{Comparison of $|\hat b(k_{in})|$ from the reduced system of equations and the time average of $|R(k_{in})|$ from the full system, Eq. (\ref{eq:partialorder}), for a network with neutral assortativity ($c=0$), $\eta_0=-2$, and $\Delta=0.1$ at $K=3$. The dynamics under these parameters were simulated in a network with $5000$ nodes, and the network was allowed to relax to a fixed point. Nodes were divided into classes according to their in-degree to calculate the time averaged effective order parameter for each class, which is shown in blue, with the error bars denoting the root mean squared time fluctuation of the order parameter for that class. The time fluctuations are due to the finite number of nodes in each class. (See text for details.)
}\label{fig:degreevariation}
\end{figure}

We now examine the transition from the PR phase to the AF phase. Microscopically, in the PR phase, almost all of the neurons are individually in a resting phase, whereas in the AF phase almost all neurons are in the spiking state. To examine the behavior at an intermediate point, we look at the fixed point for the case of $K=3$, as shown in Fig. \ref{fig:fixedpoints}(c). At this intermediate value of the coupling constant, a fraction of the neurons are in the spiking state. In particular, the nodes that begin to spike first are those which have larger in-degrees. This is demonstrated in Fig. \ref{fig:degreevariation}, in which we examine $\hat{b}(\mathbf{k})$ at the fixed point for $K=3$. Since we are looking at a network with neutral assortativity ($c=0$), Eq. (\ref{eq:splitSum}) implies that the sum only depends on the out-degree through a common multiplicative factor. Thus $\hat{b}$ is only plotted as a function of $k_{in}$. Analogously, for the fixed point of the dynamics on the full network, the range of degrees from $k_{min}$ to $k_{max}$ is divided uniformly into several intervals, and for each interval we find a partial order parameter, calculated such that the average in Eq. (\ref{eq:NetworkOrder}) is only performed over those nodes whose in-degree lie within that interval, i.e.,
\begin{equation}\label{eq:partialorder}
R(k_{in},t) = \frac{1}{||\mathcal{N}||} \sum_{j \in \mathcal{N}} e^{i \theta_j},
\end{equation}
where $\mathcal{N}$ is the set of nodes having an in-degree within one of the intervals of the range of degrees, $||\mathcal{N}||$ is the number of nodes in the set, and $k_{in}$ is the average in-degree of nodes within that set.

\begin{figure*}
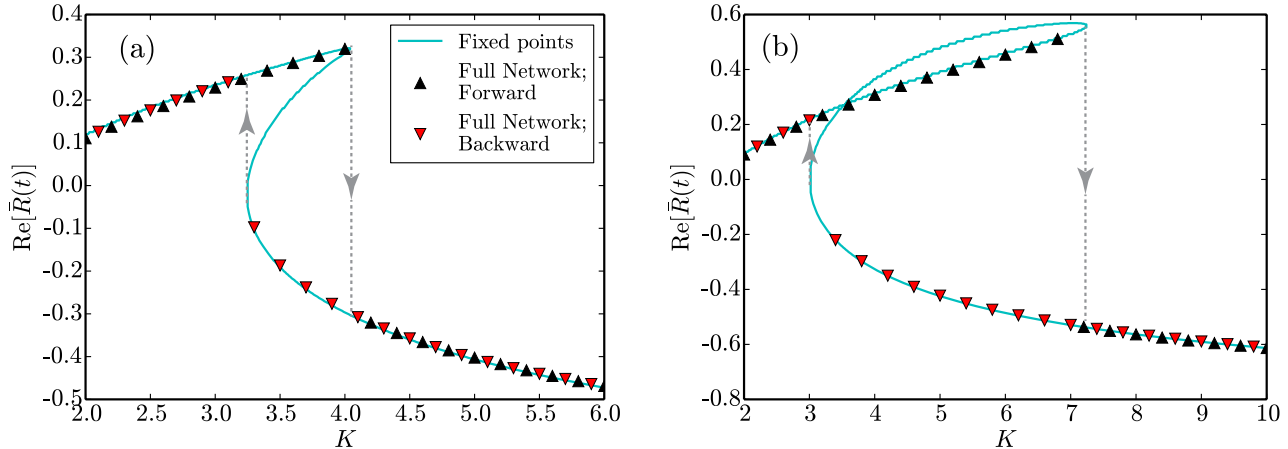

\centering
\subfigure{%
\includegraphics[width=\columnwidth]{{{Fixed_point_hysterisis_Network_N5000_c0_eta_-2_delta_0.1_real_line_fixed_points_no_errors_scale_free_a}}}
}%
\subfigure{%
\includegraphics[width=\columnwidth]{{{Fixed_point_hysterisis_Network_N5000_c0_eta_-2_delta_0.1_real_line_fixed_points_no_errors_ER_no_legend_b}}}
}%
\caption{A sweeping value of $K$ was used to observe the change in phase from the PR state to the AF state. Hysteresis was observed on the network with a scale free degree distribution (a) as well as a corresponding Erd\H{o}s-R\'{e}nyi network having the same size and the same average degree (b). For the full network, at each value of $K$ the mean of the order parameter after ignoring the transients have been marked as triangles. A close match is observed with the fixed points as computed from mean field equations directly (see text for details). Hysteresis is observed for $3.25 \lesssim K \lesssim 4$ in the scale free network (a), and is observed for $3 \lesssim K \lesssim 7.25$ in the corresponding Erd\H{o}s-R\'{e}nyi network (Note the difference in scales for the x-axis in both plots). An apparent crossing of the fixed point curve is seen in (b), which is an artifact of the non-self-intersecting $\bar{R}$ curve lying in the two dimensional complex space, which has been projected onto the real axis in this plot.
}\label{fig:hysteresis}
\end{figure*}

In addition, we find that the transition from the PR phase to the AF phase occurs via a hysteretic process mediated by saddle node bifurcations. To illustrate this, we evolved the dynamics of the full network in a step wise fashion by increasing the coupling constant $K$ in small increments of 0.2, and allowing the system to relax to an equilibrium before the next increment (Fig. \ref{fig:hysteresis}(a)). We also compare this with the analogous hysteresis curve observed for the evolution of the system dynamics on an Erd\H{o}s-R\'{e}nyi network having the same size and average degree as the scale free network being considered (Fig. \ref{fig:hysteresis}(b)). While the hysteretic region begins at around the same value of the coupling constant, $K$, for both network topologies, we find that for the case of the Erd\H{o}s-R\'{e}nyi network, which has a sharply peaked degree distribution, the range in $K$ that allows hysteresis ($3\lesssim K \lesssim 7.25$) is significantly larger than the corresponding range for the network with the scale free degree distribution ($3.25 \lesssim K \lesssim 4$). 

To compare with the simulation of the dynamics on the full network, we also calculate the fixed points of the mean field equations Eq. (\ref{eq:reducedeqn}). While the fixed points cannot be readily determined analytically, we can efficiently compute them via a numerical calculation. Setting $\partial \hat b(\mathbf{k}, t)/\partial t=0$ for the fixed points, we find that the equilibrium $\hat b(\mathbf{k})$ satisfy,

\begin{equation} \label{eq:equilB}
\hat b_\pm(\mathbf{k}) = \frac{1 \pm z(\mathbf{k})}{1 \mp z(\mathbf{k})},
\end{equation}
where

\begin{dmath} \label{eq:zExp}
i z^2(\mathbf{k}) = - \Delta + i \eta_0 + i d_n \frac{K}{\langle k \rangle} \sum_{\mathbf{k'}} P(\mathbf{k'}) a({\mathbf{k'} \rightarrow \mathbf{k}}) \\ \times  \left[ A_0 + \sum_{p=1}^n A_p (\hat b(\mathbf{k'}, t)^p + \hat b^*(\mathbf{k'}, t)^p) \right],
\end{dmath}
and the sign is chosen to ensure $|\hat b(\mathbf{k})|\leq 1$. 
Using our form of the assortativity function Eq. (\ref{eq:assort}), we may again split the above sum into two parts as in Eq. (\ref{eq:splitSum}). Thus we may rewrite Eq. (\ref{eq:zExp}) as 

\begin{equation} \label{eq:zExp2}
i z^2(\mathbf{k}) = - \Delta + i \eta_0 + i k_{in} X + i (k_{out} - \langle k \rangle) Y,
\end{equation}
where $X$ and $Y$ are given by,

\begin{dgroup} \label{eq:x/y}
\begin{dmath}X = d_n \frac{K}{N \langle k \rangle^2} \sum_{\mathbf{k'}} P(\mathbf{k'}) k_{out}'  \left[ A_0 + \sum_{p=1}^n A_p (\hat b(\mathbf{k'}, t)^p + \hat b^*(\mathbf{k'}, t)^p) \right] \end{dmath}
\begin{dmath}Y = d_n \frac{K}{N \langle k \rangle^2} \sum_{\mathbf{k'}} P(\mathbf{k'}) (k_{in}'-\langle k \rangle)  \left[ A_0 + \sum_{p=1}^n A_p (\hat b(\mathbf{k'}, t)^p + \hat b^*(\mathbf{k'}, t)^p) \right]. \end{dmath}
\end{dgroup}
These simplifications allow for efficient calculation of the system fixed points. Choosing initial values, $X_0$ and $Y_0$, we calculate the associated $z(\mathbf{k})$ and $\hat b(\mathbf{k})$ using Eq. (\ref{eq:zExp2}) and Eq. (\ref{eq:equilB}), and then recalculate  new values, $X_1$ and $Y_1$ using Eq. (\ref{eq:x/y}). For fixed points of the reduced equations $\delta X = X_1-X_0$ and $\delta Y = Y_1-Y_0$ are both zero. We calculate $\delta X$ and $\delta Y$ for several different initial values at regularly spaced intervals for $X_0$ and $Y_0$, and identify the fixed points as the points where $\delta X = \delta Y =0$. The interpolation procedure described earlier can also be applied to this calculation to further increase efficiency. For the nonassortative case ($c=0$), $Y=0$ always, so identifying the fixed points in this case only requires calculating the variation in the single parameter $X$. We use this method to evaluate the fixed points of the reduced equations for the range of $K$ over which hysteresis was observed, and find close agreement between the results of this fixed point analysis and the direct evolution of the full network (Fig. \ref{fig:hysteresis}). 

\subsection{Limit Cycles}

\begin{figure}
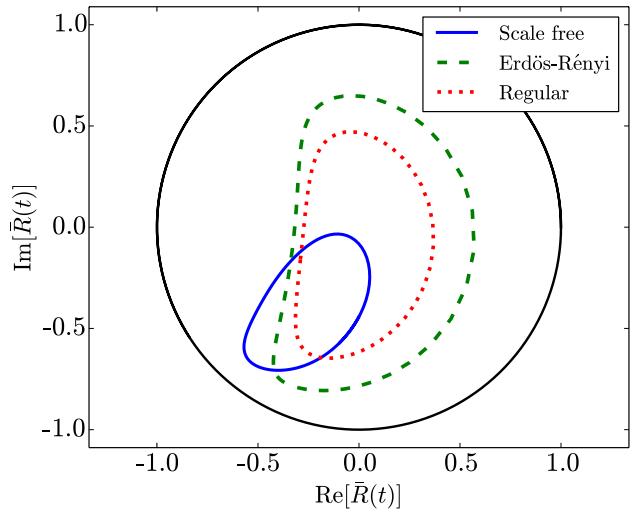

\centering
\includegraphics[width=\columnwidth]{{{Different_limit_cycles_c0_eta10.75_delta0.5_K-9_with_labels_line_dot_dashed}}}
\caption{Comparison of the limit cycle attractor for $\bar{R}(t)$ in the complex plane across varying degree distributions in a network with neutral assortativity ($c=0$) with $\eta_0=10.75$, $\Delta =0.5$ and $K=-9$. The scale free network (blue solid curve) has a degree distribution according to Eq. (\ref{eq:degdist}), the Erd\H{o}s-R\'{e}nyi network (green dashed curve) has a Poissonian degree distribution, and the regular network (red dotted curve) has a delta function degree distribution. The black circle is the unit circle $|\bar{R}|$=1
}\label{fig:limitcyclecomparison}
\end{figure}

As a representative example of limit cycles of $\bar{R}(t)$, we consider a network with neutral assortativity with excitability parameters $\eta$ distributed as a Lorentzian with mean $\eta_0=10.75$ and width $\Delta = 0.5$, and with a coupling constant $K=-9$.
In the SF phase, the order parameter goes to a limit cycle in the complex plane. In this phase, a majority of the neurons are synchronously in a spiking state. Plots for such limit cycles are shown in Fig. \ref{fig:limitcyclecomparison}, in which we plot the trajectory of the order parameter in the complex plane (after removing transients) for a network with the scale free degree distribution given in Eq. (\ref{eq:degdist}) (blue solid curve), a corresponding Erd\H{o}s-R\'{e}nyi network having a Poissonian degree distribution (green dashed curve), and a regular network having a delta function degree distribution (i.e. $P(\mathbf{k})=\delta_{k_{in}, k} \delta_{k_{out}, k}$) (red dotted curve), each having the same average degree. As seen earlier in Eq. (\ref{eq:singleEqRed}), a network with a delta function degree distribution has mean field dynamics identical to those of a fully connected network, and the corresponding limit cycle in Fig. \ref{fig:limitcyclecomparison} is identical to the limit cycle obtained at these parameters for the fully connected network by Luke et al.\cite{luke2013complete}
In comparison with the limit cycles that are observed for the case of the regular network or the Erd\H{o}s-R\'{e}nyi network, the limit cycles in networks with scale free degree distributions are diminished in size, due to the large variation in nodal behavior as a function of degree. Nodes with smaller in-degrees were observed to predominantly be in the spiking phase, with high synchronization and a larger limit cycle for the partial order parameter, whereas nodes with larger in-degrees were in the resting phase. Due to this differentiation of behavior with degree, the averaged full order parameter exhibits a limit cycle that is somewhat reduced in size when compared with the results for a fully connected network by Luke et al\cite{luke2013complete}. However, we see that the limit cycles for the Erd\H{o}s-R\'{e}nyi network are similar in shape and structure to the limit cycles obtained for the regular network, as would be expected in accordance with the discussion in Sec. \ref{sec:dimensionreduction}, since the Poissonian degree distribution for the Erd\H{o}s-R\'{e}nyi network is sharply peaked about the average degree and hence cannot admit a large variation of behavior with nodal degree. As the average degree, $\langle k \rangle$ increases, the red and green curves converge because the Poisson degree distribution appropriate for an  Erd\H{o}s-R\'{e}nyi network approaches a delta function.

\subsection{Effect of Assortativity}

\begin{figure*}
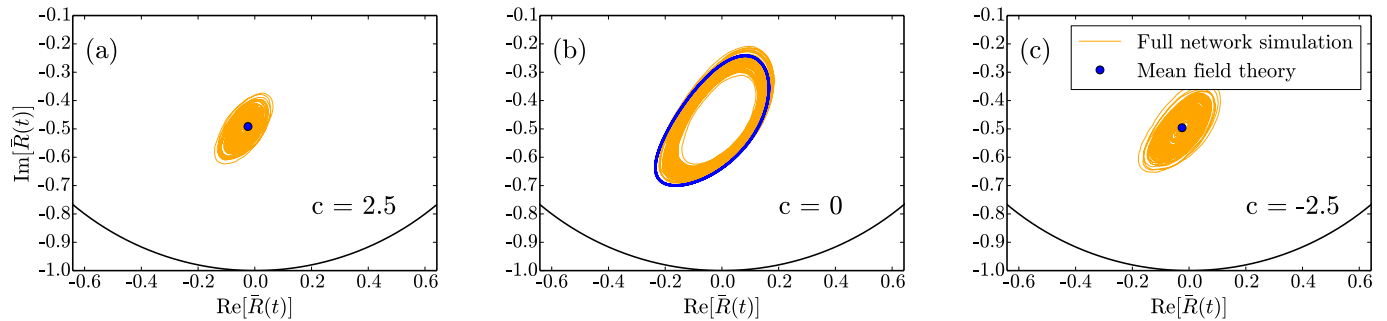

\centering
\subfigure{%
\includegraphics[height=0.525\columnwidth]{{{Limit_cycle_broken_full_network_w_mean_field_c2.5_eta_4._delta_0.5_new_R_axes_label_a}}}
}%
\subfigure{%
\includegraphics[height=0.525\columnwidth]{{{Limit_cycle_full_network_w_mean_field_c0_eta_4._delta_0.5_new_R_axes_label_b}}}
}%
\subfigure{%
\includegraphics[height=0.525\columnwidth]{{{Limit_cycle_broken_full_network_w_mean_field_c-2.5_eta_4._delta_0.5_new_R_axes_label_c}}}
}
\caption{For the parameters $\eta_0=4$, $\Delta=0.5$ and $K=-4.8$, in a network with neutral assortativity ($c=0$), the system lies in an SF state (as in (b)). Varying the assortativity in either direction ($c=\pm2.5$, corresponding to $r\approx \pm 0.198$) causes the limit cycles to be replaced by a fixed point instead (as in (a),(c)). The behavior predicted by the mean field theory is in agreement with the simulations of the full network.
}\label{fig:limitcycleassort}
\end{figure*}

We now consider the effect of assortativity on the limit cycle dynamics of the order parameter in the network. While limit cycle behavior exists in networks with neutral assortativity ($c=0$), introduction of assortativity or dissasortativity in the network can cause the limit cycle attractor to transform to a fixed point attractor (AF like state) via a Hopf bifurcation. This is demonstrated in Fig. \ref{fig:limitcycleassort}, in which we show that varying $c$ away from zero to $\pm2.5$ (corresponding to Pearson assortativity coefficients of $r\approx\pm 0.198$) is sufficient to cause the Hopf bifurcation and send the system to a fixed point attractor. The fixed points for the order parameters in these networks exhibit relatively large amounts of finite $N$ induced noise as seen from the size of the clouds surrounding the fixed point position calculated from the reduced system.

\section{Conclusion}\label{sec:conclusion}
Using a mean field approximation, in conjunction with the Ott-Antonsen ansatz, we obtained a reduced system of equations that successfully model the macroscopic order parameter dynamics of a large network of theta neurons. This reduced system of equations allows us to examine the effects of varying the network parameters and the network topology (in terms of degree distributions, as well as degree correlations) in a computationally efficient fashion. The order parameter of the network is used for describing the macroscopic behavior of the network of theta neurons, whose attractors can be of various types. In particular, we find resting states, asynchronously firing states and synchronously firing states, the first two of which appear as a fixed point for the order parameter (Fig. \ref{fig:fixedpoints}), while the third appears as a limit cycle for the order parameter (Fig. \ref{fig:limitcyclecomparison}). We also used the reduced system of equations to observe the effect of varying the assortativity in the system and demonstrated that a synchronously firing phase was only present for networks with neutral or small assortativity, and the addition of moderate amounts of assortativity or disassortativity to the network causes the system to go to an asynchronously firing state instead (Fig. \ref{fig:limitcycleassort}). Further, for networks with scale free degree distributions, we find that nodes with different values of their degrees admit a large variation of behavior (Fig. \ref{fig:degreevariation}), a phenomenon not possible in networks with all-to-all connectivity. In all cases close agreement was observed between the order parameter dynamics as predicted by the reduced system of equations (Eq. \ref{eq:reducedeqn}), and as calculated by evolution of the full system of equations Eq. (\ref{eq:network}).

\section*{Acknowledgements}

This work was supported by the Army Research Office under Grant No. W911NF-12-1-0101, and by the National Science Foundation under Grant No. PHY-1461089.

\bibliography{TREND_16_final}

\begin{thebibliography}{40}%
\makeatletter
\providecommand \@ifxundefined [1]{%
 \@ifx{#1\undefined}
}%
\providecommand \@ifnum [1]{%
 \ifnum #1\expandafter \@firstoftwo
 \else \expandafter \@secondoftwo
 \fi
}%
\providecommand \@ifx [1]{%
 \ifx #1\expandafter \@firstoftwo
 \else \expandafter \@secondoftwo
 \fi
}%
\providecommand \natexlab [1]{#1}%
\providecommand \enquote  [1]{``#1''}%
\providecommand \bibnamefont  [1]{#1}%
\providecommand \bibfnamefont [1]{#1}%
\providecommand \citenamefont [1]{#1}%
\providecommand \href@noop [0]{\@secondoftwo}%
\providecommand \href [0]{\begingroup \@sanitize@url \@href}%
\providecommand \@href[1]{\@@startlink{#1}\@@href}%
\providecommand \@@href[1]{\endgroup#1\@@endlink}%
\providecommand \@sanitize@url [0]{\catcode `\\12\catcode `\$12\catcode
  `\&12\catcode `\#12\catcode `\^12\catcode `\_12\catcode `\%12\relax}%
\providecommand \@@startlink[1]{}%
\providecommand \@@endlink[0]{}%
\providecommand \url  [0]{\begingroup\@sanitize@url \@url }%
\providecommand \@url [1]{\endgroup\@href {#1}{\urlprefix }}%
\providecommand \urlprefix  [0]{URL }%
\providecommand \Eprint [0]{\href }%
\providecommand \doibase [0]{http://dx.doi.org/}%
\providecommand \selectlanguage [0]{\@gobble}%
\providecommand \bibinfo  [0]{\@secondoftwo}%
\providecommand \bibfield  [0]{\@secondoftwo}%
\providecommand \translation [1]{[#1]}%
\providecommand \BibitemOpen [0]{}%
\providecommand \bibitemStop [0]{}%
\providecommand \bibitemNoStop [0]{.\EOS\space}%
\providecommand \EOS [0]{\spacefactor3000\relax}%
\providecommand \BibitemShut  [1]{\csname bibitem#1\endcsname}%
\let\auto@bib@innerbib\@empty
%</preamble>
\bibitem [{\citenamefont {Michaels}, \citenamefont {Matyas},\ and\
  \citenamefont {Jalife}(1987)}]{michaels1987mechanisms}%
  \BibitemOpen
  \bibfield  {author} {\bibinfo {author} {\bibfnamefont {D.~C.}\ \bibnamefont
  {Michaels}}, \bibinfo {author} {\bibfnamefont {E.~P.}\ \bibnamefont
  {Matyas}}, \ and\ \bibinfo {author} {\bibfnamefont {J.}~\bibnamefont
  {Jalife}},\ }\bibfield  {title} {\enquote {\bibinfo {title} {Mechanisms of
  sinoatrial pacemaker synchronization: a new hypothesis.}}\ }\href@noop {}
  {\bibfield  {journal} {\bibinfo  {journal} {Circulation Research}\ }\textbf
  {\bibinfo {volume} {61}},\ \bibinfo {pages} {704--714} (\bibinfo {year}
  {1987})}\BibitemShut {NoStop}%
\bibitem [{\citenamefont {Wiesenfeld}, \citenamefont {Colet},\ and\
  \citenamefont {Strogatz}(1998)}]{wiesenfeld1998frequency}%
  \BibitemOpen
  \bibfield  {author} {\bibinfo {author} {\bibfnamefont {K.}~\bibnamefont
  {Wiesenfeld}}, \bibinfo {author} {\bibfnamefont {P.}~\bibnamefont {Colet}}, \
  and\ \bibinfo {author} {\bibfnamefont {S.~H.}\ \bibnamefont {Strogatz}},\
  }\bibfield  {title} {\enquote {\bibinfo {title} {Frequency locking in
  josephson arrays: connection with the kuramoto model},}\ }\href@noop {}
  {\bibfield  {journal} {\bibinfo  {journal} {Physical Review E}\ }\textbf
  {\bibinfo {volume} {57}},\ \bibinfo {pages} {1563} (\bibinfo {year}
  {1998})}\BibitemShut {NoStop}%
\bibitem [{\citenamefont {Kiss}, \citenamefont {Zhai},\ and\ \citenamefont
  {Hudson}(2002)}]{kiss2002emerging}%
  \BibitemOpen
  \bibfield  {author} {\bibinfo {author} {\bibfnamefont {I.~Z.}\ \bibnamefont
  {Kiss}}, \bibinfo {author} {\bibfnamefont {Y.}~\bibnamefont {Zhai}}, \ and\
  \bibinfo {author} {\bibfnamefont {J.~L.}\ \bibnamefont {Hudson}},\ }\bibfield
   {title} {\enquote {\bibinfo {title} {Emerging coherence in a population of
  chemical oscillators},}\ }\href@noop {} {\bibfield  {journal} {\bibinfo
  {journal} {Science}\ }\textbf {\bibinfo {volume} {296}},\ \bibinfo {pages}
  {1676--1678} (\bibinfo {year} {2002})}\BibitemShut {NoStop}%
\bibitem [{\citenamefont {Motter}\ \emph {et~al.}(2013)\citenamefont {Motter},
  \citenamefont {Myers}, \citenamefont {Anghel},\ and\ \citenamefont
  {Nishikawa}}]{motter2013spontaneous}%
  \BibitemOpen
  \bibfield  {author} {\bibinfo {author} {\bibfnamefont {A.~E.}\ \bibnamefont
  {Motter}}, \bibinfo {author} {\bibfnamefont {S.~A.}\ \bibnamefont {Myers}},
  \bibinfo {author} {\bibfnamefont {M.}~\bibnamefont {Anghel}}, \ and\ \bibinfo
  {author} {\bibfnamefont {T.}~\bibnamefont {Nishikawa}},\ }\bibfield  {title}
  {\enquote {\bibinfo {title} {Spontaneous synchrony in power-grid networks},}\
  }\href@noop {} {\bibfield  {journal} {\bibinfo  {journal} {Nature Physics}\
  }\textbf {\bibinfo {volume} {9}},\ \bibinfo {pages} {191--197} (\bibinfo
  {year} {2013})}\BibitemShut {NoStop}%
\bibitem [{\citenamefont {Carreras}\ \emph {et~al.}(2004)\citenamefont
  {Carreras}, \citenamefont {Lynch}, \citenamefont {Dobson},\ and\
  \citenamefont {Newman}}]{carreras2004complex}%
  \BibitemOpen
  \bibfield  {author} {\bibinfo {author} {\bibfnamefont {B.~A.}\ \bibnamefont
  {Carreras}}, \bibinfo {author} {\bibfnamefont {V.~E.}\ \bibnamefont {Lynch}},
  \bibinfo {author} {\bibfnamefont {I.}~\bibnamefont {Dobson}}, \ and\ \bibinfo
  {author} {\bibfnamefont {D.~E.}\ \bibnamefont {Newman}},\ }\bibfield  {title}
  {\enquote {\bibinfo {title} {Complex dynamics of blackouts in power
  transmission systems},}\ }\href@noop {} {\bibfield  {journal} {\bibinfo
  {journal} {Chaos}\ }\textbf {\bibinfo {volume} {14}},\ \bibinfo {pages}
  {643--652} (\bibinfo {year} {2004})}\BibitemShut {NoStop}%
\bibitem [{\citenamefont {Glass}\ and\ \citenamefont
  {Kauffman}(1973)}]{glass1973logical}%
  \BibitemOpen
  \bibfield  {author} {\bibinfo {author} {\bibfnamefont {L.}~\bibnamefont
  {Glass}}\ and\ \bibinfo {author} {\bibfnamefont {S.~A.}\ \bibnamefont
  {Kauffman}},\ }\bibfield  {title} {\enquote {\bibinfo {title} {The logical
  analysis of continuous, non-linear biochemical control networks},}\
  }\href@noop {} {\bibfield  {journal} {\bibinfo  {journal} {Journal of
  Theoretical Biology}\ }\textbf {\bibinfo {volume} {39}},\ \bibinfo {pages}
  {103--129} (\bibinfo {year} {1973})}\BibitemShut {NoStop}%
\bibitem [{\citenamefont {Aldana}\ and\ \citenamefont
  {Cluzel}(2003)}]{aldana2003natural}%
  \BibitemOpen
  \bibfield  {author} {\bibinfo {author} {\bibfnamefont {M.}~\bibnamefont
  {Aldana}}\ and\ \bibinfo {author} {\bibfnamefont {P.}~\bibnamefont
  {Cluzel}},\ }\bibfield  {title} {\enquote {\bibinfo {title} {A natural class
  of robust networks},}\ }\href@noop {} {\bibfield  {journal} {\bibinfo
  {journal} {Proceedings of the National Academy of Sciences}\ }\textbf
  {\bibinfo {volume} {100}},\ \bibinfo {pages} {8710--8714} (\bibinfo {year}
  {2003})}\BibitemShut {NoStop}%
\bibitem [{\citenamefont {Luke}, \citenamefont {Barreto},\ and\ \citenamefont
  {So}(2013)}]{luke2013complete}%
  \BibitemOpen
  \bibfield  {author} {\bibinfo {author} {\bibfnamefont {T.~B.}\ \bibnamefont
  {Luke}}, \bibinfo {author} {\bibfnamefont {E.}~\bibnamefont {Barreto}}, \
  and\ \bibinfo {author} {\bibfnamefont {P.}~\bibnamefont {So}},\ }\bibfield
  {title} {\enquote {\bibinfo {title} {Complete classification of the
  macroscopic behavior of a heterogeneous network of theta neurons},}\
  }\href@noop {} {\bibfield  {journal} {\bibinfo  {journal} {Neural
  Computation}\ }\textbf {\bibinfo {volume} {25}},\ \bibinfo {pages}
  {3207--3234} (\bibinfo {year} {2013})}\BibitemShut {NoStop}%
\bibitem [{\citenamefont {Abdulrehem}\ and\ \citenamefont
  {Ott}(2009)}]{abdulrehem2009low}%
  \BibitemOpen
  \bibfield  {author} {\bibinfo {author} {\bibfnamefont {M.~M.}\ \bibnamefont
  {Abdulrehem}}\ and\ \bibinfo {author} {\bibfnamefont {E.}~\bibnamefont
  {Ott}},\ }\bibfield  {title} {\enquote {\bibinfo {title} {Low dimensional
  description of pedestrian-induced oscillation of the millennium bridge},}\
  }\href@noop {} {\bibfield  {journal} {\bibinfo  {journal} {Chaos}\ }\textbf
  {\bibinfo {volume} {19}},\ \bibinfo {pages} {013129} (\bibinfo {year}
  {2009})}\BibitemShut {NoStop}%
\bibitem [{\citenamefont {Montbri{\'o}}, \citenamefont {Paz{\'o}},\ and\
  \citenamefont {Roxin}(2015)}]{montbrio2015macroscopic}%
  \BibitemOpen
  \bibfield  {author} {\bibinfo {author} {\bibfnamefont {E.}~\bibnamefont
  {Montbri{\'o}}}, \bibinfo {author} {\bibfnamefont {D.}~\bibnamefont
  {Paz{\'o}}}, \ and\ \bibinfo {author} {\bibfnamefont {A.}~\bibnamefont
  {Roxin}},\ }\bibfield  {title} {\enquote {\bibinfo {title} {Macroscopic
  description for networks of spiking neurons},}\ }\href@noop {} {\bibfield
  {journal} {\bibinfo  {journal} {Physical Review X}\ }\textbf {\bibinfo
  {volume} {5}},\ \bibinfo {pages} {021028} (\bibinfo {year}
  {2015})}\BibitemShut {NoStop}%
\bibitem [{\citenamefont {Paz{\'o}}\ and\ \citenamefont
  {Montbri{\'o}}(2014)}]{pazo2014low}%
  \BibitemOpen
  \bibfield  {author} {\bibinfo {author} {\bibfnamefont {D.}~\bibnamefont
  {Paz{\'o}}}\ and\ \bibinfo {author} {\bibfnamefont {E.}~\bibnamefont
  {Montbri{\'o}}},\ }\bibfield  {title} {\enquote {\bibinfo {title}
  {Low-dimensional dynamics of populations of pulse-coupled oscillators},}\
  }\href@noop {} {\bibfield  {journal} {\bibinfo  {journal} {Physical Review
  X}\ }\textbf {\bibinfo {volume} {4}},\ \bibinfo {pages} {011009} (\bibinfo
  {year} {2014})}\BibitemShut {NoStop}%
\bibitem [{\citenamefont {Laing}(2014)}]{laing2014derivation}%
  \BibitemOpen
  \bibfield  {author} {\bibinfo {author} {\bibfnamefont {C.~R.}\ \bibnamefont
  {Laing}},\ }\bibfield  {title} {\enquote {\bibinfo {title} {Derivation of a
  neural field model from a network of theta neurons},}\ }\href@noop {}
  {\bibfield  {journal} {\bibinfo  {journal} {Physical Review E}\ }\textbf
  {\bibinfo {volume} {90}},\ \bibinfo {pages} {010901} (\bibinfo {year}
  {2014})}\BibitemShut {NoStop}%
\bibitem [{\citenamefont {Lu}\ \emph {et~al.}(2016)\citenamefont {Lu},
  \citenamefont {Klein-Carde{\~n}a}, \citenamefont {Lee}, \citenamefont
  {Antonsen}, \citenamefont {Girvan},\ and\ \citenamefont
  {Ott}}]{lu2016resynchronization}%
  \BibitemOpen
  \bibfield  {author} {\bibinfo {author} {\bibfnamefont {Z.}~\bibnamefont
  {Lu}}, \bibinfo {author} {\bibfnamefont {K.}~\bibnamefont
  {Klein-Carde{\~n}a}}, \bibinfo {author} {\bibfnamefont {S.}~\bibnamefont
  {Lee}}, \bibinfo {author} {\bibfnamefont {T.~M.}\ \bibnamefont {Antonsen}},
  \bibinfo {author} {\bibfnamefont {M.}~\bibnamefont {Girvan}}, \ and\ \bibinfo
  {author} {\bibfnamefont {E.}~\bibnamefont {Ott}},\ }\bibfield  {title}
  {\enquote {\bibinfo {title} {Resynchronization of circadian oscillators and
  the east-west asymmetry of jet-lag},}\ }\href@noop {} {\bibfield  {journal}
  {\bibinfo  {journal} {Chaos}\ }\textbf {\bibinfo {volume} {26}},\ \bibinfo
  {pages} {094811} (\bibinfo {year} {2016})}\BibitemShut {NoStop}%
\bibitem [{\citenamefont {Ott}\ and\ \citenamefont
  {Antonsen}(2008)}]{ott2008low}%
  \BibitemOpen
  \bibfield  {author} {\bibinfo {author} {\bibfnamefont {E.}~\bibnamefont
  {Ott}}\ and\ \bibinfo {author} {\bibfnamefont {T.~M.}\ \bibnamefont
  {Antonsen}},\ }\bibfield  {title} {\enquote {\bibinfo {title} {Low
  dimensional behavior of large systems of globally coupled oscillators},}\
  }\href@noop {} {\bibfield  {journal} {\bibinfo  {journal} {Chaos}\ }\textbf
  {\bibinfo {volume} {18}},\ \bibinfo {pages} {037113} (\bibinfo {year}
  {2008})}\BibitemShut {NoStop}%
\bibitem [{\citenamefont {Ott}\ and\ \citenamefont
  {Antonsen}(2009)}]{ott2009long}%
  \BibitemOpen
  \bibfield  {author} {\bibinfo {author} {\bibfnamefont {E.}~\bibnamefont
  {Ott}}\ and\ \bibinfo {author} {\bibfnamefont {T.~M.}\ \bibnamefont
  {Antonsen}},\ }\bibfield  {title} {\enquote {\bibinfo {title} {Long time
  evolution of phase oscillator systems},}\ }\href@noop {} {\bibfield
  {journal} {\bibinfo  {journal} {Chaos}\ }\textbf {\bibinfo {volume} {19}},\
  \bibinfo {pages} {023117} (\bibinfo {year} {2009})}\BibitemShut {NoStop}%
\bibitem [{\citenamefont {Ott}, \citenamefont {Hunt},\ and\ \citenamefont
  {Antonsen~Jr}(2011)}]{ott2011comment}%
  \BibitemOpen
  \bibfield  {author} {\bibinfo {author} {\bibfnamefont {E.}~\bibnamefont
  {Ott}}, \bibinfo {author} {\bibfnamefont {B.~R.}\ \bibnamefont {Hunt}}, \
  and\ \bibinfo {author} {\bibfnamefont {T.~M.}\ \bibnamefont {Antonsen~Jr}},\
  }\bibfield  {title} {\enquote {\bibinfo {title} {Comment on “{Long} time
  evolution of phase oscillator systems” [{Chaos} 19, 023117 (2009)]},}\
  }\href@noop {} {\bibfield  {journal} {\bibinfo  {journal} {Chaos}\ }\textbf
  {\bibinfo {volume} {21}},\ \bibinfo {pages} {025112} (\bibinfo {year}
  {2011})}\BibitemShut {NoStop}%
\bibitem [{\citenamefont {Restrepo}\ and\ \citenamefont
  {Ott}(2014)}]{restrepo2014mean}%
  \BibitemOpen
  \bibfield  {author} {\bibinfo {author} {\bibfnamefont {J.~G.}\ \bibnamefont
  {Restrepo}}\ and\ \bibinfo {author} {\bibfnamefont {E.}~\bibnamefont {Ott}},\
  }\bibfield  {title} {\enquote {\bibinfo {title} {Mean-field theory of
  assortative networks of phase oscillators},}\ }\href@noop {} {\bibfield
  {journal} {\bibinfo  {journal} {Europhysics Letters}\ }\textbf {\bibinfo
  {volume} {107}},\ \bibinfo {pages} {60006} (\bibinfo {year}
  {2014})}\BibitemShut {NoStop}%
\bibitem [{\citenamefont {Martens}\ \emph {et~al.}(2009)\citenamefont
  {Martens}, \citenamefont {Barreto}, \citenamefont {Strogatz}, \citenamefont
  {Ott}, \citenamefont {So},\ and\ \citenamefont
  {Antonsen}}]{martens2009exact}%
  \BibitemOpen
  \bibfield  {author} {\bibinfo {author} {\bibfnamefont {E.~A.}\ \bibnamefont
  {Martens}}, \bibinfo {author} {\bibfnamefont {E.}~\bibnamefont {Barreto}},
  \bibinfo {author} {\bibfnamefont {S.}~\bibnamefont {Strogatz}}, \bibinfo
  {author} {\bibfnamefont {E.}~\bibnamefont {Ott}}, \bibinfo {author}
  {\bibfnamefont {P.}~\bibnamefont {So}}, \ and\ \bibinfo {author}
  {\bibfnamefont {T.}~\bibnamefont {Antonsen}},\ }\bibfield  {title} {\enquote
  {\bibinfo {title} {Exact results for the kuramoto model with a bimodal
  frequency distribution},}\ }\href@noop {} {\bibfield  {journal} {\bibinfo
  {journal} {Physical Review E}\ }\textbf {\bibinfo {volume} {79}},\ \bibinfo
  {pages} {026204} (\bibinfo {year} {2009})}\BibitemShut {NoStop}%
\bibitem [{\citenamefont {Barlev}, \citenamefont {Antonsen},\ and\
  \citenamefont {Ott}(2011)}]{barlev2011dynamics}%
  \BibitemOpen
  \bibfield  {author} {\bibinfo {author} {\bibfnamefont {G.}~\bibnamefont
  {Barlev}}, \bibinfo {author} {\bibfnamefont {T.~M.}\ \bibnamefont
  {Antonsen}}, \ and\ \bibinfo {author} {\bibfnamefont {E.}~\bibnamefont
  {Ott}},\ }\bibfield  {title} {\enquote {\bibinfo {title} {The dynamics of
  network coupled phase oscillators: An ensemble approach},}\ }\href@noop {}
  {\bibfield  {journal} {\bibinfo  {journal} {Chaos}\ }\textbf {\bibinfo
  {volume} {21}},\ \bibinfo {pages} {025103} (\bibinfo {year}
  {2011})}\BibitemShut {NoStop}%
\bibitem [{\citenamefont {Skardal}, \citenamefont {Restrepo},\ and\
  \citenamefont {Ott}(2015)}]{skardal2015frequency}%
  \BibitemOpen
  \bibfield  {author} {\bibinfo {author} {\bibfnamefont {P.~S.}\ \bibnamefont
  {Skardal}}, \bibinfo {author} {\bibfnamefont {J.~G.}\ \bibnamefont
  {Restrepo}}, \ and\ \bibinfo {author} {\bibfnamefont {E.}~\bibnamefont
  {Ott}},\ }\bibfield  {title} {\enquote {\bibinfo {title} {Frequency
  assortativity can induce chaos in oscillator networks},}\ }\href@noop {}
  {\bibfield  {journal} {\bibinfo  {journal} {Physical Review E}\ }\textbf
  {\bibinfo {volume} {91}},\ \bibinfo {pages} {060902} (\bibinfo {year}
  {2015})}\BibitemShut {NoStop}%
\bibitem [{\citenamefont {Paz{\'o}}\ and\ \citenamefont
  {Montbri{\'o}}(2016)}]{pazo2016quasiperiodic}%
  \BibitemOpen
  \bibfield  {author} {\bibinfo {author} {\bibfnamefont {D.}~\bibnamefont
  {Paz{\'o}}}\ and\ \bibinfo {author} {\bibfnamefont {E.}~\bibnamefont
  {Montbri{\'o}}},\ }\bibfield  {title} {\enquote {\bibinfo {title} {From
  quasiperiodic partial synchronization to collective chaos in populations of
  inhibitory neurons with delay},}\ }\href@noop {} {\bibfield  {journal}
  {\bibinfo  {journal} {Physical Review Letters}\ }\textbf {\bibinfo {volume}
  {116}},\ \bibinfo {pages} {238101} (\bibinfo {year} {2016})}\BibitemShut
  {NoStop}%
\bibitem [{\citenamefont {Roulet}\ and\ \citenamefont
  {Mindlin}(2016)}]{roulet2016average}%
  \BibitemOpen
  \bibfield  {author} {\bibinfo {author} {\bibfnamefont {J.}~\bibnamefont
  {Roulet}}\ and\ \bibinfo {author} {\bibfnamefont {G.~B.}\ \bibnamefont
  {Mindlin}},\ }\bibfield  {title} {\enquote {\bibinfo {title} {Average
  activity of excitatory and inhibitory neural populations},}\ }\href@noop {}
  {\bibfield  {journal} {\bibinfo  {journal} {Chaos}\ }\textbf {\bibinfo
  {volume} {26}},\ \bibinfo {pages} {093104} (\bibinfo {year}
  {2016})}\BibitemShut {NoStop}%
\bibitem [{\citenamefont {Ermentrout}\ and\ \citenamefont
  {Kopell}(1986)}]{ermentrout1986parabolic}%
  \BibitemOpen
  \bibfield  {author} {\bibinfo {author} {\bibfnamefont {G.~B.}\ \bibnamefont
  {Ermentrout}}\ and\ \bibinfo {author} {\bibfnamefont {N.}~\bibnamefont
  {Kopell}},\ }\bibfield  {title} {\enquote {\bibinfo {title} {Parabolic
  bursting in an excitable system coupled with a slow oscillation},}\
  }\href@noop {} {\bibfield  {journal} {\bibinfo  {journal} {SIAM Journal on
  Applied Mathematics}\ }\textbf {\bibinfo {volume} {46}},\ \bibinfo {pages}
  {233--253} (\bibinfo {year} {1986})}\BibitemShut {NoStop}%
\bibitem [{\citenamefont {Ermentrout}(1996)}]{ermentrout1996type}%
  \BibitemOpen
  \bibfield  {author} {\bibinfo {author} {\bibfnamefont {B.}~\bibnamefont
  {Ermentrout}},\ }\bibfield  {title} {\enquote {\bibinfo {title} {Type i
  membranes, phase resetting curves, and synchrony},}\ }\href@noop {}
  {\bibfield  {journal} {\bibinfo  {journal} {Neural Computation}\ }\textbf
  {\bibinfo {volume} {8}},\ \bibinfo {pages} {979--1001} (\bibinfo {year}
  {1996})}\BibitemShut {NoStop}%
\bibitem [{\citenamefont {Izhikevich}(1999)}]{izhikevich1999class}%
  \BibitemOpen
  \bibfield  {author} {\bibinfo {author} {\bibfnamefont {E.~M.}\ \bibnamefont
  {Izhikevich}},\ }\bibfield  {title} {\enquote {\bibinfo {title} {Class 1
  neural excitability, conventional synapses, weakly connected networks, and
  mathematical foundations of pulse-coupled models},}\ }\href@noop {}
  {\bibfield  {journal} {\bibinfo  {journal} {IEEE Transactions on Neural
  Networks}\ }\textbf {\bibinfo {volume} {10}},\ \bibinfo {pages} {499--507}
  (\bibinfo {year} {1999})}\BibitemShut {NoStop}%
\bibitem [{\citenamefont {Hodgkin}(1948)}]{hodgkin1948local}%
  \BibitemOpen
  \bibfield  {author} {\bibinfo {author} {\bibfnamefont {A.~L.}\ \bibnamefont
  {Hodgkin}},\ }\bibfield  {title} {\enquote {\bibinfo {title} {The local
  electric changes associated with repetitive action in a non-medullated
  axon},}\ }\href@noop {} {\bibfield  {journal} {\bibinfo  {journal} {The
  Journal of Physiology}\ }\textbf {\bibinfo {volume} {107}},\ \bibinfo {pages}
  {165} (\bibinfo {year} {1948})}\BibitemShut {NoStop}%
\bibitem [{\citenamefont {B{\"o}rgers}\ and\ \citenamefont
  {Kopell}(2003)}]{borgers2003synchronization}%
  \BibitemOpen
  \bibfield  {author} {\bibinfo {author} {\bibfnamefont {C.}~\bibnamefont
  {B{\"o}rgers}}\ and\ \bibinfo {author} {\bibfnamefont {N.}~\bibnamefont
  {Kopell}},\ }\bibfield  {title} {\enquote {\bibinfo {title} {Synchronization
  in networks of excitatory and inhibitory neurons with sparse, random
  connectivity},}\ }\href@noop {} {\bibfield  {journal} {\bibinfo  {journal}
  {Neural computation}\ }\textbf {\bibinfo {volume} {15}},\ \bibinfo {pages}
  {509--538} (\bibinfo {year} {2003})}\BibitemShut {NoStop}%
\bibitem [{\citenamefont {Newman}(2002)}]{newman2002assortative}%
  \BibitemOpen
  \bibfield  {author} {\bibinfo {author} {\bibfnamefont {M.~E.}\ \bibnamefont
  {Newman}},\ }\bibfield  {title} {\enquote {\bibinfo {title} {Assortative
  mixing in networks},}\ }\href@noop {} {\bibfield  {journal} {\bibinfo
  {journal} {Physical review letters}\ }\textbf {\bibinfo {volume} {89}},\
  \bibinfo {pages} {208701} (\bibinfo {year} {2002})}\BibitemShut {NoStop}%
\bibitem [{\citenamefont {Hagmann}\ \emph {et~al.}(2008)\citenamefont
  {Hagmann}, \citenamefont {Cammoun}, \citenamefont {Gigandet}, \citenamefont
  {Meuli}, \citenamefont {Honey}, \citenamefont {Wedeen},\ and\ \citenamefont
  {Sporns}}]{hagmann2008mapping}%
  \BibitemOpen
  \bibfield  {author} {\bibinfo {author} {\bibfnamefont {P.}~\bibnamefont
  {Hagmann}}, \bibinfo {author} {\bibfnamefont {L.}~\bibnamefont {Cammoun}},
  \bibinfo {author} {\bibfnamefont {X.}~\bibnamefont {Gigandet}}, \bibinfo
  {author} {\bibfnamefont {R.}~\bibnamefont {Meuli}}, \bibinfo {author}
  {\bibfnamefont {C.~J.}\ \bibnamefont {Honey}}, \bibinfo {author}
  {\bibfnamefont {V.~J.}\ \bibnamefont {Wedeen}}, \ and\ \bibinfo {author}
  {\bibfnamefont {O.}~\bibnamefont {Sporns}},\ }\bibfield  {title} {\enquote
  {\bibinfo {title} {Mapping the structural core of human cerebral cortex},}\
  }\href@noop {} {\bibfield  {journal} {\bibinfo  {journal} {PLoS Biology}\
  }\textbf {\bibinfo {volume} {6}},\ \bibinfo {pages} {e159} (\bibinfo {year}
  {2008})}\BibitemShut {NoStop}%
\bibitem [{\citenamefont {Bialonski}\ and\ \citenamefont
  {Lehnertz}(2013)}]{bialonski2013assortative}%
  \BibitemOpen
  \bibfield  {author} {\bibinfo {author} {\bibfnamefont {S.}~\bibnamefont
  {Bialonski}}\ and\ \bibinfo {author} {\bibfnamefont {K.}~\bibnamefont
  {Lehnertz}},\ }\bibfield  {title} {\enquote {\bibinfo {title} {Assortative
  mixing in functional brain networks during epileptic seizures},}\ }\href@noop
  {} {\bibfield  {journal} {\bibinfo  {journal} {Chaos}\ }\textbf {\bibinfo
  {volume} {23}},\ \bibinfo {pages} {033139} (\bibinfo {year}
  {2013})}\BibitemShut {NoStop}%
\bibitem [{\citenamefont {Barzegaran}\ \emph {et~al.}(2012)\citenamefont
  {Barzegaran}, \citenamefont {Joudaki}, \citenamefont {Jalili}, \citenamefont
  {Rossetti}, \citenamefont {Frackowiak},\ and\ \citenamefont
  {Knyazeva}}]{barzegaran2012properties}%
  \BibitemOpen
  \bibfield  {author} {\bibinfo {author} {\bibfnamefont {E.}~\bibnamefont
  {Barzegaran}}, \bibinfo {author} {\bibfnamefont {A.}~\bibnamefont {Joudaki}},
  \bibinfo {author} {\bibfnamefont {M.}~\bibnamefont {Jalili}}, \bibinfo
  {author} {\bibfnamefont {A.~O.}\ \bibnamefont {Rossetti}}, \bibinfo {author}
  {\bibfnamefont {R.~S.}\ \bibnamefont {Frackowiak}}, \ and\ \bibinfo {author}
  {\bibfnamefont {M.~G.}\ \bibnamefont {Knyazeva}},\ }\bibfield  {title}
  {\enquote {\bibinfo {title} {Properties of functional brain networks
  correlate with frequency of psychogenic non-epileptic seizures},}\
  }\href@noop {} {\bibfield  {journal} {\bibinfo  {journal} {Frontiers in Human
  Neuroscience}\ }\textbf {\bibinfo {volume} {6}} (\bibinfo {year}
  {2012})}\BibitemShut {NoStop}%
\bibitem [{\citenamefont {de~Haan}\ \emph {et~al.}(2009)\citenamefont
  {de~Haan}, \citenamefont {Pijnenburg}, \citenamefont {Strijers},
  \citenamefont {van~der Made}, \citenamefont {van~der Flier}, \citenamefont
  {Scheltens},\ and\ \citenamefont {Stam}}]{de2009functional}%
  \BibitemOpen
  \bibfield  {author} {\bibinfo {author} {\bibfnamefont {W.}~\bibnamefont
  {de~Haan}}, \bibinfo {author} {\bibfnamefont {Y.~A.}\ \bibnamefont
  {Pijnenburg}}, \bibinfo {author} {\bibfnamefont {R.~L.}\ \bibnamefont
  {Strijers}}, \bibinfo {author} {\bibfnamefont {Y.}~\bibnamefont {van~der
  Made}}, \bibinfo {author} {\bibfnamefont {W.~M.}\ \bibnamefont {van~der
  Flier}}, \bibinfo {author} {\bibfnamefont {P.}~\bibnamefont {Scheltens}}, \
  and\ \bibinfo {author} {\bibfnamefont {C.~J.}\ \bibnamefont {Stam}},\
  }\bibfield  {title} {\enquote {\bibinfo {title} {Functional neural network
  analysis in frontotemporal dementia and alzheimer's disease using eeg and
  graph theory},}\ }\href@noop {} {\bibfield  {journal} {\bibinfo  {journal}
  {BMC Neuroscience}\ }\textbf {\bibinfo {volume} {10}},\ \bibinfo {pages} {1}
  (\bibinfo {year} {2009})}\BibitemShut {NoStop}%
\bibitem [{\citenamefont {Teller}\ \emph {et~al.}(2014)\citenamefont {Teller},
  \citenamefont {Granell}, \citenamefont {De~Domenico}, \citenamefont
  {Soriano}, \citenamefont {G{\'o}mez},\ and\ \citenamefont
  {Arenas}}]{teller2014emergence}%
  \BibitemOpen
  \bibfield  {author} {\bibinfo {author} {\bibfnamefont {S.}~\bibnamefont
  {Teller}}, \bibinfo {author} {\bibfnamefont {C.}~\bibnamefont {Granell}},
  \bibinfo {author} {\bibfnamefont {M.}~\bibnamefont {De~Domenico}}, \bibinfo
  {author} {\bibfnamefont {J.}~\bibnamefont {Soriano}}, \bibinfo {author}
  {\bibfnamefont {S.}~\bibnamefont {G{\'o}mez}}, \ and\ \bibinfo {author}
  {\bibfnamefont {A.}~\bibnamefont {Arenas}},\ }\bibfield  {title} {\enquote
  {\bibinfo {title} {Emergence of assortative mixing between clusters of
  cultured neurons},}\ }\href@noop {} {\bibfield  {journal} {\bibinfo
  {journal} {PLoS Computational Biology}\ }\textbf {\bibinfo {volume} {10}},\
  \bibinfo {pages} {e1003796} (\bibinfo {year} {2014})}\BibitemShut {NoStop}%
\bibitem [{Note1()}]{Note1}%
  \BibitemOpen
  \bibinfo {note} {Some authors, such as Restrepo and Ott\cite
  {restrepo2014mean} define the order parameter differently so as to be
  weighted with the out-degree at each node, i.e., $R(t) = \sum _{i=1}^N \sum
  _{j=1}^N A_{ij} e^{i \theta _j} / \left (\sum _{i=1}^N \sum _{j=1}^N
  A_{ij}\right )$}\BibitemShut {NoStop}%
\bibitem [{\citenamefont {Foster}\ \emph {et~al.}(2010)\citenamefont {Foster},
  \citenamefont {Foster}, \citenamefont {Grassberger},\ and\ \citenamefont
  {Paczuski}}]{foster2010edge}%
  \BibitemOpen
  \bibfield  {author} {\bibinfo {author} {\bibfnamefont {J.~G.}\ \bibnamefont
  {Foster}}, \bibinfo {author} {\bibfnamefont {D.~V.}\ \bibnamefont {Foster}},
  \bibinfo {author} {\bibfnamefont {P.}~\bibnamefont {Grassberger}}, \ and\
  \bibinfo {author} {\bibfnamefont {M.}~\bibnamefont {Paczuski}},\ }\bibfield
  {title} {\enquote {\bibinfo {title} {Edge direction and the structure of
  networks},}\ }\href@noop {} {\bibfield  {journal} {\bibinfo  {journal}
  {Proceedings of the National Academy of Sciences}\ }\textbf {\bibinfo
  {volume} {107}},\ \bibinfo {pages} {10815--10820} (\bibinfo {year}
  {2010})}\BibitemShut {NoStop}%
\bibitem [{Note2()}]{Note2}%
  \BibitemOpen
  \bibinfo {note} {For another, often useful, definition of a coefficient
  quantitatively characterizing the assortativity or disassortativity of a
  network see Ref.\cite {restrepo2007approximating}}\BibitemShut {NoStop}%
\bibitem [{Note3()}]{Note3}%
  \BibitemOpen
  \bibinfo {note} {This definition of asynchronous spiking is consistent with
  remarks by other authors \cite {abbott1993asynchronous, hansel2001existence},
  wherein asynchronous states have been defined as states in which at each
  neuron the term coupling it to the other neurons in the network is
  independent of time, as is observed in the cases of fixed
  points.}\BibitemShut {Stop}%
\bibitem [{\citenamefont {Restrepo}, \citenamefont {Ott},\ and\ \citenamefont
  {Hunt}(2007)}]{restrepo2007approximating}%
  \BibitemOpen
  \bibfield  {author} {\bibinfo {author} {\bibfnamefont {J.~G.}\ \bibnamefont
  {Restrepo}}, \bibinfo {author} {\bibfnamefont {E.}~\bibnamefont {Ott}}, \
  and\ \bibinfo {author} {\bibfnamefont {B.~R.}\ \bibnamefont {Hunt}},\
  }\bibfield  {title} {\enquote {\bibinfo {title} {Approximating the largest
  eigenvalue of network adjacency matrices},}\ }\href@noop {} {\bibfield
  {journal} {\bibinfo  {journal} {Physical Review E}\ }\textbf {\bibinfo
  {volume} {76}},\ \bibinfo {pages} {056119} (\bibinfo {year}
  {2007})}\BibitemShut {NoStop}%
\bibitem [{\citenamefont {Abbott}\ and\ \citenamefont {van
  Vreeswijk}(1993)}]{abbott1993asynchronous}%
  \BibitemOpen
  \bibfield  {author} {\bibinfo {author} {\bibfnamefont {L.}~\bibnamefont
  {Abbott}}\ and\ \bibinfo {author} {\bibfnamefont {C.}~\bibnamefont {van
  Vreeswijk}},\ }\bibfield  {title} {\enquote {\bibinfo {title} {Asynchronous
  states in networks of pulse-coupled oscillators},}\ }\href@noop {} {\bibfield
   {journal} {\bibinfo  {journal} {Physical Review E}\ }\textbf {\bibinfo
  {volume} {48}},\ \bibinfo {pages} {1483} (\bibinfo {year}
  {1993})}\BibitemShut {NoStop}%
\bibitem [{\citenamefont {Hansel}\ and\ \citenamefont
  {Mato}(2001)}]{hansel2001existence}%
  \BibitemOpen
  \bibfield  {author} {\bibinfo {author} {\bibfnamefont {D.}~\bibnamefont
  {Hansel}}\ and\ \bibinfo {author} {\bibfnamefont {G.}~\bibnamefont {Mato}},\
  }\bibfield  {title} {\enquote {\bibinfo {title} {Existence and stability of
  persistent states in large neuronal networks},}\ }\href@noop {} {\bibfield
  {journal} {\bibinfo  {journal} {Physical Review Letters}\ }\textbf {\bibinfo
  {volume} {86}},\ \bibinfo {pages} {4175} (\bibinfo {year}
  {2001})}\BibitemShut {NoStop}%
\end{thebibliography}%

\end{document}